\documentclass[conf]{new-aiaa}
\usepackage[utf8]{inputenc}

\usepackage{graphicx}
\usepackage{amsmath}
\usepackage[version=4]{mhchem}
\usepackage{siunitx}
\usepackage{longtable,tabularx}
\usepackage{cite}
\usepackage{xcolor}
\title{Attitude Determination and Control of GPS Satellites:
Stabilization, Orbital Insertion, and Operational Control
Mechanisms}

\author{Oliullah Samir\footnote{PhD Student, Department of Mechanical and Aerospace Engineering, The University of Texas at Arlington, Arlington, TX 76019.}}
\affil{The University of Texas at Arlington, Arlington, TX 76019}

\begin{document}

\maketitle

\begin{abstract}
Global Positioning System (GPS) satellites are essential for providing accurate navigation and timing information worldwide. Operating in medium Earth orbit (MEO), these satellites must maintain precise Earth-pointing attitudes to transmit signals effectively. This paper presents a comprehensive review of the operational dynamics, attitude determination
and control systems (ADCS), and orbital insertion techniques for GPS satellites. We explore the integration of sensors and actuators, control algorithms, stabilization strategies, and the launch procedures required to deploy these satellites. Key equations related to orbital mechanics and attitude control are discussed, and references to recent technical literature are included.

\end{abstract}

\section{Nomenclature}

{\renewcommand\arraystretch{1.0}
\noindent\begin{longtable*}{@{}l @{\quad=\quad} l@{}}
$\mu$  & Earth’s gravitational parameter \\
$a$ &    Semi Major Axis \\
$v_e$& effective exhaust velocity of the rocket \\
$R_e$ & Radius of Earth \\
$I$ &  satellite’s inertia matrix \\
$\omega$ & satellite’s Angular Velocity Vector \\
$T$ & External Torque \\
$\hat{x}_k$ & Estimated State \\
$K_k$   & Kalman Gain \\
$z_k$  & Measured Output \\
$H$  & Observation Matrix
\end{longtable*}}

\section{Introduction}
\lettrine{T}he Global Positioning System (GPS) is a satellite-based radio navigation system developed by the U.S. Department of Defense, initially conceptualized in the 1970s and declared fully operational in 1995 \citep{parkinson1996gps}. The constellation is composed of at least 24 active satellites arranged in six orbital planes inclined at approximately 55 degrees, with each satellite orbiting at an altitude of about 20,200 km in Medium Earth Orbit (MEO) \citep{kaplan2006understanding}. These orbital parameters allow each satellite to complete an orbit in roughly 11 hours and 58 minutes, thereby enabling continuous and repeatable coverage patterns globally \citep{hofmann2001global}.

\begin{figure}[hbt!]
\centering
\includegraphics[width=.5\textwidth]{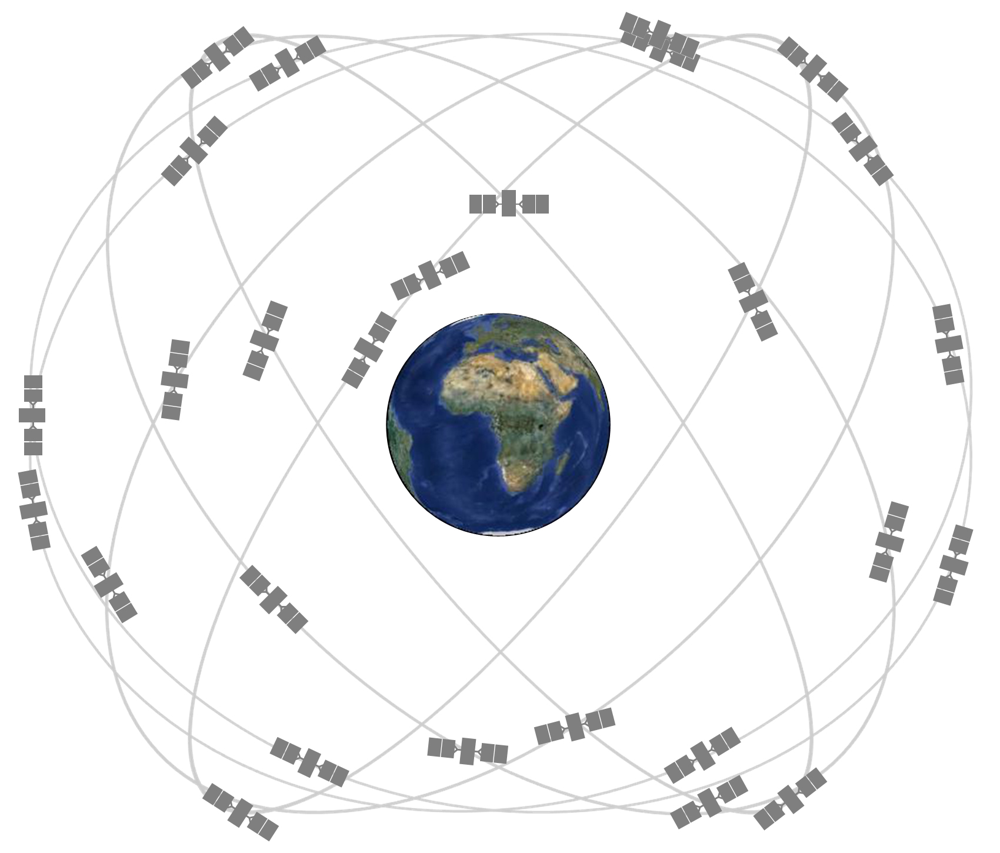}
\caption{Simplified diagram of the GPS satellite constellation (Expandable 24-slot configuration). The satellites occupy six orbital planes (gray circular orbits) inclined ~55° to the equator, with four primary satellites in each plane. This geometry ensures global coverage with at least four satellites in view from any location on Earth.}
\label{fig:Figure_1}
\end{figure}

Figure~\ref{fig:Figure_1} illustrates the GPS constellation’s six-plane configuration, with four satellites in each plane providing overlapping coverage. This arrangement ensures that at least four satellites are above the horizon from any point on Earth at all times, which is the minimum needed for precise trilateration. Each satellite transmits a unique signal containing its ephemeris data, almanac information, and a highly accurate timestamp synchronized to onboard atomic clocks \citep{misra2006global}. The system is managed through three major segments: the space segment (satellites), the control segment (ground monitoring and updates), and the user segment (civilian and military receivers).

The utility of GPS extends far beyond navigation. It is vital for time synchronization in power grids, mobile communication networks, and financial systems. It also plays a pivotal role in aviation, marine navigation, autonomous vehicles, precision agriculture, surveying, and Earth science applications. In atmospheric science, GPS signals are used in radio occultation to infer temperature, pressure, and moisture content in the atmosphere .

Beyond Earth, the principles underlying GPS have inspired the development of planetary navigation concepts. While Mars currently lacks a dedicated satellite navigation infrastructure, future Mars exploration missions envision deploying orbiters with GPS-like capabilities to support autonomous rover navigation and reduce reliance on Earth-based tracking \citep{9764737}. This is especially relevant for planetary rovers, such as NASA’s Perseverance and Curiosity, which currently use a combination of inertial measurement units (IMUs), visual odometry, and orbital imagery for localization\citep{Zaman2024}\citep{wristsms}. Integrating satellite-based positioning systems in future Mars missions could dramatically improve rover autonomy, enable precise landing site targeting, and support large-scale coordinated surface exploration.

Engineering principles used in GPS design—such as dimensional analysis, scaled-model experimentation, and high-fidelity computational simulation—find parallels in other engineering disciplines. For example, \citep{turbine} \citep{sina3d} \citep{Albaruni2025} demonstrated how experimental data from scaled Archimedes spiral wind turbine models, combined with computational fluid dynamics (CFD) simulations, can accurately predict the performance of full-scale systems. Similar methodologies are used in GPS antenna design, signal coverage prediction, and orbital dynamics validation before full deployment.

Multiphysics modeling also plays a key role in GPS system engineering, especially when analyzing structural responses to dynamic loads. \citep{dynamicsmaruf} \citep{sina3d2} applied a non-smooth multiphysics modeling approach to investigate tibiofemoral joint contact forces under high-frequency impact loading. This approach, though developed for biomechanics, parallels the modeling techniques used to predict satellite structural responses to launch vibrations, mechanical shocks, and in-orbit disturbances.

Material science advancements are equally vital for GPS satellite reliability. The selection of thin film coatings affects resistance to UV radiation, thermal cycling, and micrometeoroid impacts. \citep{ABDULLAH2025114117} \citep{motayed4480455heat}reviewed the prospects and challenges of thin film coating technologies, highlighting their potential to enhance spacecraft surface durability and operational longevity in the harsh space environment.

This paper provides an in-depth technical review of GPS satellite design and operation, focusing on launch dynamics, orbital insertion strategies, and the Attitude Determination and Control Systems (ADCS) essential for maintaining accurate orientation in space. The integration of advanced sensors and control algorithms ensures that each satellite maintains proper Earth-pointing orientation for reliable signal transmission. 

\section{Orbital Insertion of GPS Satellites}
GPS satellites are deployed into Medium Earth Orbit (MEO) at an altitude of 20,200 km, with a semi-major axis of 26,560 km, an inclination of 55 degrees, and a period of 43,200 seconds (half a sidereal day). The orbital velocity is calculated as \citep{curtisfundamentals}:

\begin{equation}
v = \sqrt{\frac{\mu}{a}} = 3.87 \, \text{km/s}
\end{equation}

where \(\mu = 3.986 \times 10^{14} \, \text{m}^3/\text{s}^2\) (Earth’s gravitational parameter), and \(a = 26,560 \, \text{km}\).

\subsection{Launch Vehicle Dynamics}
GPS satellites, weighing 2,000–4,000 kg (e.g., GPS III at 3,880 kg), are launched using vehicles like the Delta IV, Atlas V, or SpaceX Falcon 9. The Falcon 9, known for its reusable first stage, reduces launch costs. The launch sequence includes:

\begin{enumerate}
    \item \textbf{Boost Phase}: The first stage, powered by Merlin engines with a specific impulse of 282 seconds at sea level, delivers a \(\Delta v\) of approximately 3 km/s. The nine Merlin 1D engines produce a total thrust of 7,607 kN, enabling the rocket to escape Earth’s lower atmosphere \citep{sutton2016rocket}.
    \item \textbf{Upper Stage}: The second stage, with a single Merlin 1D Vacuum engine (specific impulse 348 seconds), performs burns to reach a transfer orbit. It delivers a \(\Delta v\) of about 7 km/s, culminating in a total \(\Delta v\) of 10 km/s for MEO insertion \citep{isakowitz2004international}.
    \item \textbf{Orbit Insertion}: The satellite is released into MEO after the upper stage executes its final burn, typically at the apogee of the transfer orbit.
\end{enumerate}

The rocket equation governs the total \(\Delta v\):

\begin{equation}
\Delta v = v_e \ln \left( \frac{m_0}{m_f} \right)
\end{equation}

For a Falcon 9 launching a GPS III satellite, the mass ratio \(m_0/m_f \approx 20\), yielding \(\Delta v \approx 13 \, \text{km/s}\), sufficient to overcome gravity losses (2 km/s) and atmospheric drag \citep{turner2009rocket}. The upper stage burn is precisely timed to minimize fuel consumption, with a typical burn duration of 300 seconds, delivering a thrust of 934 kN \citep{sutton2016rocket}.

\subsection{Hohmann Transfer Mechanics}
The transfer to MEO employs a Hohmann transfer from a 200 km low Earth orbit (LEO) parking orbit (\(r_1 = 6,578 \, \text{km}\)) to the target MEO (\(r_2 = 26,560 \, \text{km}\)) \citep{curtisfundamentals}. A Hohmann transfer is an efficient two-impulse orbital maneuver for coplanar, circular orbits. The first impulse (burn) raises the satellite onto an elliptical transfer orbit, and the second impulse at the elliptical orbit’s apogee circularizes the orbit at the new altitude. During the first burn, the spacecraft’s velocity is increased to enter the elliptical transfer orbit (raising the apogee to MEO altitude). During the second burn at apogee, the velocity is increased again to raise the perigee and circularize at 20,200 km. The total \(\Delta v\) required for the Hohmann transfer can be computed from orbital mechanics\citep{curtisfundamentals}. The \(\Delta v\) for each maneuver is:
\begin{equation}
\Delta v_1 = \sqrt{\frac{\mu}{r_1}} \left( \sqrt{\frac{2r_2}{r_1 + r_2}} - 1 \right) \approx 2.45 \, \text{km/s}
\end{equation}

\begin{equation}
\Delta v_2 = \sqrt{\frac{\mu}{r_2}} \left( 1 - \sqrt{\frac{2r_1}{r_1 + r_2}} \right) \approx 1.47 \, \text{km/s}
\end{equation}

The total \(\Delta v\) is approximately 3.92 km/s. The transfer orbit’s semi-major axis is:

\begin{equation}
a_{\text{transfer}} = \frac{r_1 + r_2}{2} = 16,569 \, \text{km}
\end{equation}

with an eccentricity of 0.73, and the transfer time is half the orbital period, approximately 5.3 hours \citep{curtisfundamentals}. The inclination of the transfer orbit is initially set to match the 55-degree inclination of the target MEO, requiring a plane change during launch to align with the desired orbital plane \citep{griffin1980principles}.

\begin{figure}[hbt!]
\centering
\includegraphics[width=.5\textwidth]{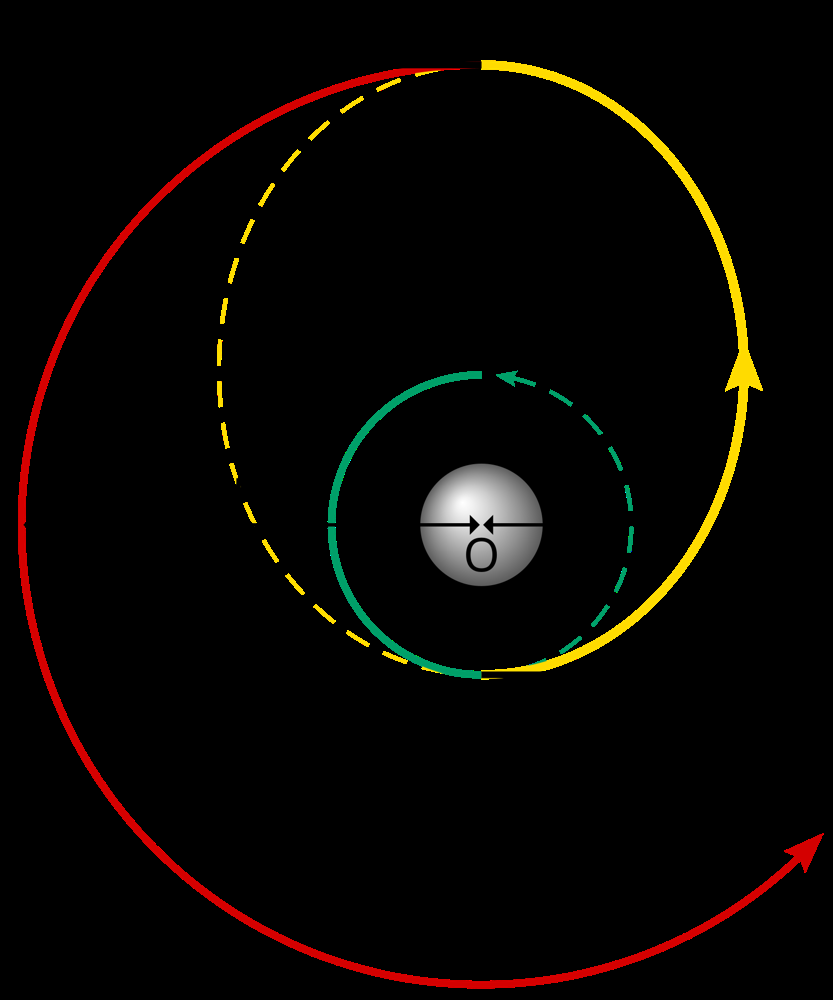}
\caption{Hohmann Transfer Trajectory.}
\label{fig:Figure_2}
\end{figure}

Figure~\ref{fig:Figure_2} illustrates a Hohmann transfer orbit (yellow trajectory) from a lower circular parking orbit (green) to a higher circular orbit (red). Two engine impulses (burns) are applied: one to insert the satellite into the elliptical transfer orbit, and a second to circularize at the MEO altitude. This is the most fuel-efficient two-impulse transfer for coplanar orbits. Perturbations during the transfer, such as Earth’s oblateness (J2 effect), solar radiation pressure, and lunar-solar gravitational influences, can cause deviations. The J2 perturbation introduces a nodal regression rate of:

\begin{equation}
\dot{\Omega} = -\frac{3}{2} \frac{J_2 R_e^2}{\sqrt{\mu a^{7/2}}} \cos i \approx -0.03 \, \text{deg/day}
\end{equation}

where \(J_2 = 1.0826 \times 10^{-3}\), \(R_e = 6,378 \, \text{km}\), and \(i = 55^\circ\). This effect is small over the 5.3-hour transfer but requires monitoring \citep{vallado2013fundamentals}.

\subsection{Upper Stage and Onboard Propulsion}
The upper stage performs the final burn at the apogee of the transfer orbit to circularize the orbit at 20,200 km. The burn adjusts the velocity to match the circular orbit speed of 3.87 km/s, requiring precise timing within a 10-second window to achieve an insertion accuracy of \(\pm 1 \, \text{km}\) in semi-major axis \citep{isakowitz2004international}. The upper stage uses a restartable engine, allowing multiple burns to correct for any initial errors during the ascent \citep{sutton2016rocket}.

After separation, the GPS satellite’s onboard propulsion system, typically hydrazine thrusters with a specific impulse of 220 seconds, performs fine adjustments. The thrust force is:

\begin{equation}
F = \dot{m} \cdot v_e
\end{equation}

where \(\dot{m} \approx 0.01 \, \text{kg/s}\), and \(v_e \approx 2,200 \, \text{m/s}\). These thrusters deliver a \(\Delta v\) of 10–20 m/s to correct residual errors in semi-major axis, eccentricity, and inclination, ensuring the satellite settles into its operational orbit within a \(\pm 2^\circ\) longitude window \citep{soop1994handbook} \citep{goebel2008fundamentals}.

\subsection{Accuracy Requirements and Error Mitigation}
GPS satellites require high insertion accuracy to ensure proper constellation geometry and minimize station-keeping fuel consumption. The target tolerances are \citep{kelso2007gps}:
- Semi-major axis: \(\pm 1 \, \text{km}\)
- Eccentricity: \(\leq 0.005\)
- Inclination: \(\pm 0.1^\circ\)

Errors in the upper stage burn, such as thrust misalignment (0.1-degree error causing a 5 m/s velocity error) or timing inaccuracies, can lead to deviations. These are mitigated using:\\
\\
-  \textbf{Pre-Launch Calibration}: The launch vehicle’s inertial navigation system is calibrated to an accuracy of 0.01 degrees \citep{wertz2011space}.\\
-  \textbf{Real-Time Telemetry}: Ground stations monitor the upper stage trajectory, providing corrective commands during the burn \citep{vallado2013fundamentals}.\\
-  \textbf{Onboard Autonomy}: The satellite uses star trackers (1 arcsecond accuracy) and GPS receivers to autonomously refine its orbit post-separation \citep{sharma2018optical}.
\subsection{Post-Insertion Verification}
After insertion, the satellite’s orbit is verified using ground-based tracking stations, part of the GPS Control Segment. The MCS in Colorado Springs uses radar and optical tracking to determine the initial orbit with a position accuracy of 10 meters and velocity accuracy of 0.1 m/s \citep{misra2006global}. The orbit determination process employs a Kalman filter:

\begin{equation}
\ddot{\mathbf{r}} = -\frac{\mu}{r^3} \mathbf{r} + \mathbf{a}_{\text{pert}}
\end{equation}

where \(\mathbf{a}_{\text{pert}}\) includes perturbations like solar radiation pressure (\(10^{-7} \, \text{m}/\text{s}^2\)) and lunar gravity (\(10^{-6} \, \text{m}/\text{s}^2\)). If discrepancies are detected, the satellite performs corrective maneuvers, typically requiring a \(\Delta v\) of 5–10 m/s, to align with the target orbit \citep{chao2005orbit}. This verification phase lasts 24–48 hours, after which the satellite begins its operational checkout, including antenna deployment and signal testing \citep{kelso2007gps}.

\subsection{Challenges in Orbital Insertion}
The insertion process faces several challenges:
\begin{itemize}
    \item \textbf{Launch Windows}: Constrained to a 15-minute window daily due to orbital plane alignment \citep{griffin1980principles}.
    \item \textbf{Vibration and Shock}: Random vibrations (10 g RMS) and acoustic loads (140 dB) are mitigated using composite structures \citep{wijker2004spacecraft}.
    \item \textbf{Thermal Stresses}: Managed by thermal blankets (emissivity 0.05) and pre-launch conditioning to 20°C \citep{karhunen2015thermal}.
    \item \textbf{Separation Mechanism}: A spring-loaded system with a separation velocity of 0.5 m/s ensures clean release \citep{sarafin1995spacecraft}.
    \item \textbf{Space Debris Risk}: The trajectory avoids debris belts, and the satellite includes shielding (1 mm aluminum), with a 1\% risk of micrometeoroid penetration over 15 years \citep{klinkrad2006space}.
\end{itemize}

\section{Satellite Control}
\subsection{Attitude Control Systems}
Once in orbit, a GPS satellite must orient itself correctly: its body must be stabilized such that the Earth-facing side (with the antenna array) is directed toward Earth, and solar panels are oriented toward the Sun. GPS satellites typically employ a combination of three-axis stabilization using reaction wheels and magnetic torquers, with thrusters available for momentum management and coarse maneuvers. This active Attitude Determination and Control Subsystem (ADCS) maintains the satellite’s attitude within tight pointing requirements (to keep antenna boresight toward Earth and maintain signal strength).
\begin{itemize}
    \item \textbf{Actuators}: Reaction wheels provide precise torque control by exchanging angular momentum between the satellite's body and the spinning flywheels \citep{brown2002spacecraft}. Magnetorquers interact with the Earth’s magnetic field to generate control torques without expending fuel, making them ideal for momentum dumping or long-duration missions \citep{wertz2011space}.

    The governing equation for satellite attitude dynamics is Euler’s rotational equation:
\begin{equation}
\mathbf{I} \dot{\boldsymbol{\omega}} + \boldsymbol{\omega} \times \mathbf{I} \boldsymbol{\omega} = \boldsymbol{T}
\end{equation}
    This equation is used in the control algorithms to predict how the satellite responds to applied torques.
    \item \textbf{Sensors and Attitude Determination}: Accurate attitude knowledge is achieved by combining multiple sensors. GPS satellites commonly use coarse Sun sensors and Earth sensors for initial attitude acquisition (to find the Sun and Earth direction upon startup) and star trackers for precise attitude determination during normal operations. Star trackers are optical devices that recognize star patterns to output the spacecraft’s attitude quaternion with arc-second accuracy. In addition, three-axis gyroscopes (often fiber-optic or MEMS gyros in modern satellites) provide angular rate measurements. The gyros drift over time, but their high-rate data bridged between star tracker updates. Magnetometers may also be included to sense Earth’s magnetic field for coarse attitude info and to assist magnetorquer control. The integration of these sensors’ data is typically done through an Extended Kalman Filter (EKF) or similar state estimator \citep{psiaki2005attitude} \citep{samiulsensor2025} \citep{awal2024model} \citep{sinaest2025}. The EKF propagates the satellite’s attitude state (often parameterized by quaternions or Euler angles and gyro bias terms) using gyro measurements and periodically corrects this estimate with absolute attitude measurements from star trackers and Sun sensors.The EKF is generated by the following equation
\begin{equation}
    \hat{\mathbf{x}}_{k|k} = \hat{\mathbf{x}}_{k|k-1} + K_k (\mathbf{z}_k - H \hat{\mathbf{x}}_{k|k-1})
\end{equation}
    \item \textbf{Control Algorithms}: Once attitude is determined, the control laws command the reaction wheels and magnetorquers to null any pointing error. GPS satellites typically employ either a proportional-derivative (PD) controller, a proportional-integral-derivative (PID) controller, or more advanced state-space controllers like Linear Quadratic Regulators (LQR) for attitude control. A PD controller uses the attitude error (e.g., difference between current and desired quaternion/angles) and the rotation rate to apply a corrective torque that is proportional to the error and its derivative. This provides damping and smooth correction with minimal steady-state error for stable pointing. PID control can add an integral term to eliminate steady-state bias (for instance, countering a constant disturbance torque such as solar radiation pressure on an asymmetrical antenna panel). LQR controllers take a state-space approach, defining a cost function (e.g., weighted sum of angle error and control effort) and solving the Riccati equation to yield an optimal feedback gain matrix. An LQR can account for cross-coupling between axes and efficiently handle multi-axis maneuvers, at the cost of requiring a good model of the satellite’s inertia and dynamics. In practice, many GPS satellites use a form of PD control for nominal pointing due to its robustness and simplicity, and augment it with wheel momentum management and safe-mode logic \citep{wertz1978attitude}.
\end{itemize}
\begin{figure}[hbt!]
\centering
\includegraphics[width=.5\textwidth]{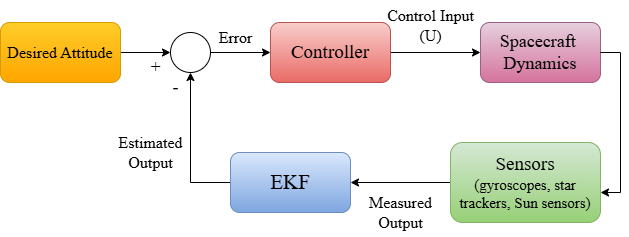}
\caption{Satellite Control Loop.}
\label{fig:Figure_3}
\end{figure}
Figure~\ref{fig:Figure_3} illustrates How the overall satellite control systems work including the EKF estimation. Overall, the attitude control system in GPS satellites is designed to be highly autonomous and fault-tolerant. If the spacecraft senses an anomaly (for example, loss of attitude knowledge or a wheel failure), it can enter a safe-hold mode. In safe mode, the satellite typically points its solar panels toward the Sun (to maintain power) and might use simple magnetic or thruster-based control to slow any rotation, awaiting ground commands. Such contingencies are part of the ADCS design to ensure the satellite can recover from sensor or actuator failures. Modern GPS III satellites have improved cross-strapping of sensors and actuators, and more advanced algorithms, to enhance reliability of the ADCS. The continued evolution of control laws (including experiments with adaptive or machine-learning-based controllers) aims to reduce ground intervention and improve pointing accuracy under all conditions.

\subsection{Station-Keeping}
In addition to attitude stabilization, GPS satellites require station-keeping to counteract gravitational perturbations and maintain their assigned orbital slot within the constellation. Station-keeping maneuvers are typically performed using onboard chemical thrusters to correct drift caused by Earth’s oblateness (effect), lunar and solar gravitational influences, and solar radiation pressure \citep{fortescue2011spacecraft, montenbruck2001satellite}.
Station-keeping maintains position within a \(\pm 2^\circ\) longitude window:

\begin{equation}
\Delta v = \sqrt{\mu \left( \frac{2}{r} - \frac{1}{a} \right)} - \sqrt{\frac{\mu}{a}}
\end{equation}

For a 1 km adjustment, \(\Delta v \approx 1.5 \, \text{m/s}\). Inclination corrections (0.02 degrees/year drift) require \(\Delta v \approx 0.5 \, \text{m/s}\), with a fuel budget of 50 kg of hydrazine over 15 years\citep{chao2005orbit}.
For long-term missions, autonomous control algorithms are implemented to reduce ground intervention. These algorithms monitor onboard telemetry and execute small corrections without ground command, improving satellite operational longevity and reliability \citep{guelman1993autonomous}.
\section{Conclusion}
GPS satellites require precise attitude determination and control to ensure accurate navigation services. The integration of robust sensors, actuators, and control laws allows them to maintain Earth-pointing orientation and solar alignment in the dynamic environment of MEO. Proper orbital insertion using transfer orbits and launch vehicles is essential to initiate their mission. The continued evolution of ADCS and launch systems ensures the reliability of future satellite navigation infrastructure.
\section*{Acknowledgments}
The author would like to acknowledge the first two figures illustrating the GPS satellite constellation and the Hohmann transfer orbit were created with the aid of AI-based image generation tools. The final interpretations, writings, and analyses are the sole responsibility of the author.


\bibliography{References}

\end{document}